\input  phyzzx
\input epsf
\overfullrule=0pt
\hsize=6.5truein
\vsize=9.0truein
\voffset=-0.1truein
\hoffset=-0.1truein

%
%

\def\IC{{\ \hbox{{\rm I}\kern-.6em\hbox{\bf C}}}}
\def\IR{{\hbox{{\rm I}\kern-.2em\hbox{\rm R}}}}
\def\IZ{{\hbox{{\rm Z}\kern-.4em\hbox{\rm Z}}}}

\def\sIR{{\hbox{{\sevenrm I}\kern-.2em\hbox{\sevenrm R}}}}

\def\sym{super Yang Mills theory}

\def\n.{{N \over 2}}
\def\bt{boundary theory}
%
%
\hyphenation{Min-kow-ski}
\rightline{SU-ITP-98-39, IASSNS-HEP-98-44}
\rightline{hep-th/9805114}
\rightline{May, 1998}

\vfill

%
%
\title{The Holographic Bound in Anti-de Sitter Space}

\vfill

%
%
\author{L. Susskind$^1$}

\vfill

\address{$^1$Department of Physics,
Stanford University\break Stanford, CA
94305-4060}
\vfill
\author{E. Witten$^2$}

\vfill
\address{$^2$School of Natural Sciences, Institute For Advanced Study
Princeton NJ 08540}


\vfill

%
%
The correspondence between string theory in
Anti-de Sitter space and super Yang Mills theory is an
example of the Holographic principle according to which a quantum
theory with gravity
must be describable by a boundary theory. However,
arguments given so far are incomplete because, while 
the bulk theory has been related
to a boundary theory, the holographic bound saying that the boundary
theory has only one bit of information per Planck area
has not been justified.
We show here that this bound is the physical interpretation of one
of the unusual aspects of the correspondence between Anti-de Sitter
space and the boundary conformal field theory, which is that
infrared effects in the bulk theory are reflected as ultraviolet
effects in the boundary theory.

\vfill\endpage

%
%



\REF\holo{L. Susskind,    The World as a Hologram,  hep-th/9409089}


\REF\hoof{ C.R. Stephens, G.
       't Hooft and B.F. Whiting,  Black Hole Evaporation Without
Information Loss, gr-qc/9310006.}
 \REF\juan{J. M. Maldacena, The Large N Limit of Superconformal Field
Theories and Supergravity, hep-th/9711200 .}
\REF\gkp{S. S. Gubser, I. R. Klebanov, and A. M. Polyakov, ``Gauge Theory
Correlators From Noncritical String Theory,'' hep-th/9802109.}
\REF\witt{ E. Witten,  Anti De Sitter Space And Holography, 
hep-th/9802150.}
\REF\gubk{S. S. Gubser and I. R. Klebanov, ``Absorption By Branes
And Schwinger Terms In The World Volume Theory,'' Phys. Lett. {\bf B413}
(1997) 41.}
 \REF\gubs{       S.S. Gubser, I. R. Klebanov, A.W. Peet, Entropy and
Temperature of Black 3-Branes, hep-th/9602135.}
\REF\hawking{S. W. Hawking and D. Page, ``Thermodynamics Of Black
Holes In Anti-de Sitter Space,'' Commun. Math. Phys. {\bf 87} (1983) 577.}


%
%

%
%
\chapter{Introduction}

According to the holographic [\holo,\hoof] hypothesis, a macroscopic region
of space and
everything inside it can be represented by a boundary
theory living on the boundary of the region. Furthermore, the \bt \ should
not contain more than one degree of freedom per
Planck area. More precisely, the number of distinct quantum states should
not exceed $\exp {A\over 4G_D}$. Here $A$
represents the $d-1$ dimensional area in a $d+1 =D$   dimensional spacetime
and $G$ is the gravitational constant in $D$
dimensions. One might imagine that the boundary theory is cutoff or
discrete so that the information density is bounded.

Some recent support for this view has come from the study of Type IIB string
theory on the background $AdS_5 \times S^5$, with a
characteristic  radius $R$ for both factors and $N$ units of five-form
flux on $S^5$. In particular,
this theory appears to be dual to $3+1$ dimensional
$U(N)$ \sym \ with 16 real supercharges [\juan]. The \sym \
lives on the boundary of the $AdS$ space.  It has been possible [\gkp,\witt]
to describe a precise recipe expressing correlation functions of the boundary
theory in terms of calculations performed in the bulk.

Though this equivalence of a bulk theory with gravity to a boundary
theory without gravity is an important part of the holographic
hypothesis, another important aspect has not yet been addressed in
the literature.  This is the holographic bound on the information density
of the boundary theory: it should have only a finite number of degrees
of freedom per Planck area.
In fact, the counting requires some care, 
because in the usual form of the correspondence the entropy and area
are both infinite.  The entropy of the boundary theory is infinite,
because this theory  
  is a conformal field theory which  has degrees
of freedom at arbitrarily small scale.  On the other hand, the boundary
of the $AdS$ space has infinite area. There is no contradiction, just
a question of whether there is a natural way to regulate and compare
these two infinities.

We will see that the essence of the matter has to do with the following
fact about the correspondence between $AdS$ space and the conformal
field theory on the boundary: infrared effects in $AdS$ space correspond
to ultraviolet effects in the boundary theory.  This shows up in many
aspects of the correspondence between these two types of theory,
going back to brane scattering computations [\gubk] from which the
equivalence was first guessed.
For example [\witt], 
relevant, marginal, and irrelevant perturbations of the boundary
conformal field theory (which are perturbations that vanish, remain
constant, or diverge as one goes to the ultraviolet), are mapped 
to perturbations of the $AdS$ space that vanish, remain constant,
or diverge as one goes to spatial infinity, that is to the infrared.
As a variant of this (see the discussion of gravity in section 2.4 of
[\witt]), ultraviolet divergences of the boundary theory in coupling
to a background gravitational field, and the resulting conformal anomaly,
are derived from an infrared divergence in computing the total volume
of the interior.  We will call this relation the I.R. - U.V. connection.
The contribution of the present paper is to show
that the I.R. - U.V. connection is the key to the information
bound that is an important part of the holographic hypothesis.

We begin by explaining why  the I.R. - U.V. connection is natural
given the geometry of $AdS$ space.  Then we explain
why it is the key to the ``information bound'' in the holographic hypothesis,
and conclude with some general remarks on holography in $AdS$ space.

%
%
\chapter{$AdS$ Space}

There are many ways to present $AdS$ space. For our purposes we find it
particularly convenient to represent it as a product
of a unit four dimensional spatial ball  with an infinite time axis
\footnote*{$AdS$ space is sometimes assumed to
have a periodic time. In this paper we work on the covering space for which
$- \infty <t < + \infty$. Unlike some other
coordinates the coordinates we use cover the entire $AdS$ in a single valued
manner.}.  The metric has the form
$$
dS^2 = R^2\left[ {4 dx^i dx^i \over (1-r^2)^2} - dt^2{1+r^2 \over 1-r^2} 
\right]
\eqn\twoone
$$
where $i=1,..,4$ and $r^2 = x^i x^i$. The $AdS$ space is the ball $r<1$. The
boundary conformal theory lives on
the sphere $r=1$.

We will use the notation $x$ for points in the bulk of
the space and $X$ for points on the boundary. Our conventions will be as
follows.
When discussing the bulk theory, distances will mean proper distance as
defined by \twoone. On the other hand when discussing
the surface theory distances on the unit sphere will be defined to be
dimensionless and given by the metric \twoone \ without
the factor $R^2$. Similarly concepts such as temperature in the boundary
theory will be defined to be dimensionless. The
dimensions can be restored with the appropriate factors of $R$.

The correspondence between the bulk supergravity in the ball and the
surface \sym \ requires a relationship between R, the radius
of the $AdS$, and $N$, the size of the gauge group [\juan]:
$$
R=l_s (g_s N)^{1/4}
\eqn\twotwo
$$
Here $g_{s} , l_{s}$ are the string coupling constant and string length
scale.

The duality between the two theories is expressed in terms of correlators
on the boundary. In
particular, supergravity field correlators $G(x_1)G(x_2)$ of various kinds
should be equal to super Yang Mills correlators
$Y(X_1)Y(X_2)$ when the points $x$ are brought to the boundary points $X$.
Let us consider the behavior of these correlators in
some more detail. We will assume that the fields $Y$ are dimensionless.
This can always be arranged by introducing an
arbitrary regulator mass scale $\mu$. Because the
\sym
\ is a conformal field theory, the operator products
$Y(X_1)Y(X_2)$ should have the form
$$
Y(X_1)Y(X_2) = \mu^{-p} |X_1 -X_2|^{-p}+\dots
\eqn\twothree
$$
for some $p$
when the coordinate distance $|X_1 -X_2|$ on the unit sphere tends to zero.

In what follows we will regulate the area of the boundary by replacing it
with a sphere just inside the boundary at $r=1-\delta$,
where $\delta$ is a number much smaller than $1$. The resulting area of the
sphere is
$$
A\approx {R^3 \over \delta^3}
\eqn\twofour
$$
Now consider the geodesic  distance between two points $X_1,X_2$ on the
regulated sphere (that is, the length of that part of a geodesic connecting
$X_1$ and $X_2$ that lies at $r<1-\delta$). 
One easily finds that it is of order
$\log(|X_1-X_2|/\delta)=\log |X_1-X_2| -\log \delta$. 
(In fact, the $\delta$ dependence comes from the divergence of the length
of the geodesic as $\delta\to 0$; the dependence on $|X_1-X_2|$ then follows
on dimensional grounds.)
A typical propagator for a particle of mass $m$ in the
bulk theory therefore has the behavior
$$
\Delta (X_1,X_2) = \exp  {m [ log \delta -log |X_1-X_2| ]} ={\delta ^m
\over  |X_1-X_2|^m }
\eqn\twofive
$$
for $|X_1-X_2|>> \delta $. For distances of order $\delta$ or smaller the
power law is not correct. The effective theory on the regulated sphere is
modified
 at small distances.

Comparing \twothree \ and \twofive, we first of all see how it is possible
for massive propagators in the bulk theory to be
represented by power laws in the conformal theory. We also see that the
infrared regulator $\delta$ in the bulk theory
 also plays the role of an
ultraviolet 
regulator in the boundary \sym. Thus we see that regulating the large
boundary
area is represented by a short distance regulator in
the \sym. We refer to this as the I.R.-U.V. connection.

Another example of this I.R. - U.V. connection can be seen by considering a
string of the bulk theory which is stretched across a
diameter of the ball. Its energy is easily computed as an integral along the
string. One finds that it is linearly divergent near
the boundary. The meaning of this is as follows. A string ending in the
boundary is represented as a point charge in the \sym. The
linearly divergent energy is the infinite self-energy of a point charge in
$3+1$-dimensional gauge theory. However, if we
regulate the sphere then the linearly divergent energy becomes proportional
to $\delta^{-1}$, which is exactly what we would
expect from a U.V. cutoff in the \sym .

%
%
\chapter{Information and Cutoffs}

In the last section, we saw that regulating the boundary area is equivalent
to U.V. regulating the \sym.  We will now make some
intuitively plausible assumptions about the information storage capacity of
a cutoff field theory. Introducing a cutoff in
field theory can be viewed as replacing the space that the field theory
lives in by discrete cells of the cutoff size. In this
case we replace the regulated sphere by cells of coordinate size $\delta$
($\delta$
is dimensionless).
We assume that each independent quantum field is
replaced by a single degree of freedom in each cell. We also assume that
each degree of freedom is capable of storing a single
bit of information.

The assumption can also be stated in terms of a limitation of the allowable
states of the cutoff quantum field theory. For
example we can limit the energy density carried by a single quantum field
to be no larger than $\delta ^{-4}$. Another
possibility is to limit the local temperature to be less than
$\delta^{-1}$. All of these give the same answer, so we will simply
say that there is one degree of freedom per cell per field degree of
freedom. We are now ready to count.

The total number of cells making up the sphere is of order $\delta^{-3}$,
and the number of field degrees of freedom in a $U(N)$
theory is of order $N^2$. Thus the number of degrees of freedom is
$$
N_{dof} = {N^2 \over \delta^3}
\eqn\threeone
$$

Using \twofour \ we can write \threeone \ as
$$
N_{dof}={A N^2 \over R^3}
\eqn\threetwo
$$
Now using \twotwo \ we get
$$
N_{dof}={A R^5 \over l_s^8 g_s^2}
\eqn\threethree
$$
Finally we recognize $l_s^8 g_s^2 R^{-5}$ as the 5 dimensional
gravitational constant so that we find
$$
N_{dof}=A/G_5
\eqn\threefour
$$
Apart from the numerical constant $1/4$, this is the desired result.

As a further illustration of the connection between the U.V. cutoff of the \sym
\ and the large area regulator in the bulk supergravity theory, let us
consider a
thermal state of the \sym \ at temperature $T$. Such a thermal ensemble
must represent
an $AdS$ Schwarzschild black hole centered at the center of the ball [\witt].
The \sym
\ equation  of state  [\gubs] has the form
$$
S=N^2 T_{ym}^3
\eqn\threefive
$$
where  $T_{ym}$ is the dimensionless temperature of the \sym .
Using \twotwo \  and $T= T_{ym}/R$ we find
$$
(TR)^3 ={Sg_s^2 l_s^8 \over R^8}
\eqn\threesix
$$
which in fact is the correct connection between temperature and entropy for
AdS black holes. If we now use
the usual Bekenstein formula for the connection between area and black hole
entropy we find
$$
T_{ym}^3= A/R^3
\eqn\threeseven
$$
Now let us suppose that the  \sym \ is regulated by saying that the maximum
value of $T_{ym}$ is $1/ \delta$.
This corresponds to a black hole of maximum area given by
$$
A_{max} = R^3 / \delta^3
\eqn\threeeight
$$
This is the area of a sphere $r=1-\delta$ which is indeed the largest
sphere that is allowed in the regulated theory.
Thus we again see how a U.V. cutoff in the boundary theory is connected to
an I.R. cutoff in the bulk theory.

\chapter{General Remarks On Holography}

We will close with some comments about the nature of the mapping
from the bulk to the boundary theory. In a conformally invariant
theory, one can not only move things around in position; one can
also rescale them.  For example, if the theory has  closed strings,
then it must have strings of every size.  
The scale size of the string
or other object is the degree of freedom which becomes the coordinate
perpendicular to the boundary.\foot{This becomes clear 
if one represents $AdS$ space by the metric $ds^2={(1/(x^0)^2)}(
(dx_0)^2+\sum_{i=1}^4
(dx^i)^2)$, where $x^i$ are the boundary coordinates, and $x^0$ controls
the distance from the boundary.  A dilation of the boundary is
generated by the isometry $x^i\to \lambda x^i$,
$x^0\to \lambda x^0$ of $AdS$ space; so rescalings of an object change
the distance from the boundary.}
In particular, the  I.R. - U.V.
connection suggests that  small (big)  in the \sym \ sense means near
the boundary (center) of the ball. This is easy to see directly in the $AdS$
space. Take a region of size $a$ near  near the center  of the ball.
Now transport it by a  conformal transformation (that is, by an element
of the $AdS$ symmetry group $SO(4,2)$ or $SO(5,1)$)
to a point at a coordinate distance $\delta$ from the
boundary. The region will be shrunk to a
coordinate size $\delta a$. Thus it seems that scale size gets transmuted
into a spatial dimension. The one thing which is far from obvious is why
the \sym \ should behave locally in the scale size.

To summarize, we have shown that when suitably regulated, the \sym \
boundary theory
provides a true holographic description including the bound of one bit per
Planck area.
To quantify just how strange and  wonderful this is, consider the number of
degrees per
unit  volume.  If we had a holographic theory in an asymptotically
flat space with spatial dimension $d$, we would simply reason that
the number of degrees of freedom within a volume $V$ (granted an ultraviolet
cutoff at the Planck scale) would be proportional to $V$ in an ordinary
local field theory, but to $A=V^{(d-1)/d}$ in a holographic theory.
Simply because
 $A/V\to 0$ for $V\to \infty$, holography entails a drastic reduction
of the number of degrees of freedom.  The same argument cannot
be made quite as easily in a world of negative cosmological constant,
since if one keeps fixed the radius $R$ of curvature, then $A$ and $V$
are proportional to one another for $V\to\infty$.  In fact, the relation
between them is asymptotically $V=AR$ in $AdS$ space, or $V=AR^6$ for
$AdS_5\times S^5$.

However, we can see the dramatic effects of holography if we vary
the $N$ of the boundary $U(N)$ gauge theory and thus let $R$ vary.
The relation $V=AR^6$ becomes using \threethree\
$$
{N_{dof}\over V}={A  \over R l_s^8 g_s^2}
\eqn\threenine
$$
As $R$ becomes large the number of degrees of freedom per unit volume tends
to zero.
Nevertheless, the theory is capable of describing string theory with a
length scale $l_{st}$ in the bulk space that is independent of $R$.

Another way to make the same point is to consider the high temperature
behavior of the entropy.  A local field in $AdS_n$ has an entropy
of order $T^{n-1}$ at high temperature, as noted in [\hawking].
This is the standard high temperature scaling in $n-1$ space dimensions;
a salient point is that the coefficient of $T^{n-1}$ is finite
because of a red-shifting of the local temperature pointed out in [\hawking].
In $AdS_5$, this would give $T^4$ for the entropy of a local field at
high temperatures.
In $AdS_5\times S^5$, a local field would have high temperature entropy
of order $T^9$.  But the boundary conformal field theory has a high
temperature entropy that is only of order $T^3$ for large $T$, showing
that a holographic theory has much lower entropy than a local field
theory in the same spacetime.

Given any theory in $AdS$ spacetime, whether it contains gravity or not,
one can define local operators in a putative boundary theory by
considering, just as in [\gkp,\witt], the boundary behavior of perturbations
in the bulk theory.  In this way, given any $AdS$ theory, one can construct
candidate correlation functions of a  boundary theory.   
Is the bulk theory equivalent to the
boundary theory that has these correlation functions?  The answer is
evidently ``no,'' if the bulk theory does not have gravity, since the
entropies disagree at high $T$, or perhaps more intuitively, since the
bulk theory (being a local field theory without gravity)
has ordinary local observables in the interior as well as local
observables on the boundary.  It is quite possible, however,
that when the bulk theory has gravity, the answer is always ``yes.''
This hypothesis is a sharpened form of the holographic hypothesis,
for the case of theories with a negative cosmological constant.

It remains to ask whether one can build a similarly sharpened holographic
hypothesis for theories with zero (or even positive)
cosmological constant.  The answer will require some new ideas,
since Minkowski space (or de Sitter space)
 has no obvious close analog of the ``boundary at spatial infinity''
by which holography is realized when the cosmological constant is negative.
\chapter{Acknowledgments}
We would like to thank Nathan Seiberg,  Steve Shenker, and Arvind Rajaraman for
valuable discussions and for raising helpful questions.
L.S. acknowledges the support of the NSF under Grant No. PHY - 9219345
and E.W. acknowledges the support of NSF Grant PHY - 9513835.

\refout
\end